\documentclass[graybox]{svmult}
\usepackage{mathptmx}       
\usepackage{helvet}         
\usepackage{courier}        
\usepackage{type1cm}        
\usepackage{makeidx}         
\usepackage{graphicx}        
\usepackage{multicol}        
\usepackage[bottom]{footmisc}

\makeindex             
\begin{document}
\title*{Extremely Compact Massive Galaxies at $1.7<z<3$}
\author{Fernando Buitrago, Ignacio Trujillo, Christopher J. Conselice}
\institute{Fernando Buitrago \at School of Physics and Astronomy, University of Nottingham, NG7 2RD, UK, \email{ppxfb@nottingham.ac.uk}
\and Ignacio Trujillo \at Instituto de Astrof\'{i}sica de Canarias, V\'{i}a L\'{a}ctea s/n 38200, La Laguna, Tenerife, Spain, \email{trujillo@iac.es}
\and Christopher J. Conselice \at School of Physics and Astronomy, University of Nottingham, NG7 2RD, UK, \email{conselice@nottingham.ac.uk}}
%
%
\maketitle
\vspace{-3cm}
\abstract*{We measure and analyse the sizes of 82 massive ($M\geq10^{11} M_{\odot}$) galaxies at $1.7\leq z\leq 3$ utilizing deep HST NICMOS data taken in the GOODS North and South fields. Our sample provides the first statistical study of massive galaxy sizes at $z>2$. We split our sample into disk--like (S\'{e}rsic index $n\leq 2$) and spheroid--like (S\'{e}rsic index $n>2$) galaxies, and find that at a given stellar mass, disk--like galaxies at $z\sim2.3$ are a factor of $2.6\pm0.3$ smaller than present day equal mass systems, and spheroid--like galaxies at the same redshift are $4.3\pm0.7$ times smaller than comparatively massive elliptical galaxies today. We furthermore show that the stellar mass densities of very massive galaxies at $z\sim2.5$ are similar to present--day globular clusters with values $\sim2\times10^{10 }M_{\odot} kpc^{-3}$}
\abstract{We measure and analyse the sizes of 82 massive ($M\geq10^{11} M_{\odot}$) galaxies at $1.7\leq z\leq 3$ utilizing deep HST NICMOS data taken in the GOODS North and South fields. Our sample provides the first statistical study of massive galaxy sizes at $z>2$. We split our sample into disk--like (S\'{e}rsic index $n\leq 2$) and spheroid--like (S\'{e}rsic index $n>2$) galaxies, and find that at a given stellar mass, disk--like galaxies at $z\sim2.3$ are a factor of $2.6\pm0.3$ smaller than present day equal mass systems, and spheroid--like galaxies at the same redshift are $4.3\pm0.7$ times smaller than comparatively massive elliptical galaxies today. We furthermore show that the stellar mass densities of very massive galaxies at $z\sim2.5$ are similar to present--day globular clusters with values $\sim2\times10^{10 }M_{\odot} kpc^{-3}$}
\section{Introduction}
\label{sec:1}
One of the most exciting discoveries in extragalactic astronomy in the last few
years is that massive ($M\geq10^{11} M_{\odot}$) galaxies at $z>1$ were
extremely compact (~\cite{1}; ~\cite{2}). Since these objects are not
found in the local Universe (~\cite{3}) it is
clear that significant growth in the sizes of these galaxies has occurred
during cosmic history.

How these compact galaxies form and evolve is not understand.
Models suggest that at very early times galaxies contain 
large amounts of cold
gas, resulting in efficient starbursts (e.g., ~\cite{4}).
As star formation occurs, the gas in these galaxies becomes heated and removed due to 
various feedback processes, leading to reduced
star formation rates, creating compact and massive remnants that are very poor in gas.
This way, ``dry mergers'' are expected to be the dominant mechanism for size and
stellar mass growth for these objects ~\cite{5}.
At $z>$2, however, our knowledge of the size evolution of the most massive
objects is much more scarce. There are only a few attempts to explore this
issue  using small
samples of massive galaxies at z$\sim$2.5 (e.g. ~\cite{6}).
With the aim of substantially increase our knowledge of the
size evolution of massive galaxies in at this redshift we analyzed a sample
of 82 massive galaxies ~\cite{7} that reveal a continual decrease in
galaxy size at $z>2$.
\section{Our Sample: GOODS NICMOS Survey (GNS)}
\label{sec:2}
The GOODS NICMOS Survey (NGS) (PI C.J. Conselice) is a large HST NICMOS--3 camera program of 60 pointings  centered around massive galaxies at $z = 1.7-3$ at 3 orbits depth, for a total of 180 orbits in the F160W (H) band. Each tile ($52"\times52", 0.203"/pix$) was observed in six exposures combined to produce images with a pixel scale of $0.1"$, and a Point Spread Function (PSF) of $\sim0.3"$ Full Width Half Maximum (FWHM). We optimized our pointings to obtain as many high-mass galaxies as possible. These galaxies consist of Distant Red Galaxies, IEROs and BzK galaxies (~\cite{7}).  Within our NICMOS fields we find a total of 82 galaxies with masses larger than $M\geq10^{11} h_{70}^{-2} M_{\odot}$  with photometric and spectroscopic redshifts in the range $1.7\leq z\leq 3$. In addition to these data, and to allow a comparison with the sizes obtained in the H--band, we measure, whenever possible, the sizes of the same galaxies using the z--band (F850LP, 5 orbits/image) HST ACS data. The z-band data is drizzled to a scale 0.03"/pix and has a PSF FWHM of $\sim0.1''$. Limiting magnitudes reached are $H=26.8(5\sigma)$ and $z=27$ ($10\sigma$ in a $0.2”$ aperture) ~\cite{8}.
\section{Results}
\label{sec:3}
Masses and photometric redshifts of our galaxy sample, although we have some spectroscopic ones, are derived from standard multi-color stellar population fitting techniques (e.g., ~\cite{9}), using filters BVRIizJHK. In particular, our stellar masses are calculated by assuming a Chabrier Initial Mass Function (IMF) and producing model Spectral Energy Distributions (SEDs) constructed from Bruzual \& Charlot ~\cite{10} stellar populations synthesis models. Galaxy sizes were measured using the GALFIT code ~\cite{11}. We check in addition the structural parameters using ACS data ~\cite{8}, utilizing simulations ~\cite{12} on our ability to recover and measure these systems.

\begin{figure}[h]
\centering
%
\includegraphics[scale=.55]{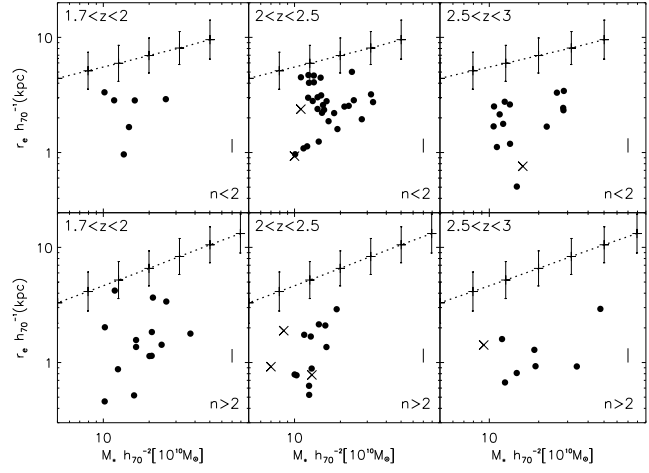}
\caption{Stellar mass-size distribution for our sample. Overplotted are the mean and the $1 \sigma$ dispersion of the distribution of the S\'{e}rsic half-light radius
of SDSS as a function of stellar mass ~\cite{3} and the crosses are the galaxies from ~\cite{6} whose masses have been converted to our Chabrier IMF. For clarity, individual error bars are not shown for our data, but a typical size error bar is shown in the right side of each bin. This mean size relative error is 0.04" which is 0.32 kpc at z=2.5. Uncertainties in the stellar masses are $\sim0.2$ dex.}
\label{fig:1}       
\end{figure}

We present the stellar mass-size relation of our sample in Fig. ~\ref{fig:1}. The first redshift bin allows a comparison with previous work ~\cite{13}. Overplotted on each panel is the local value of the half--light radii, and its dispersion, at a given stellar mass ~\cite{3} from Sloan Digital Sky Survey. Remarkably, none of our galaxies at $z>1.7$ fall in the mean distribution of the local relation, and only three would match if the masses were overestimated by a factor of two.

\begin{figure}[b]
\centering
%
\includegraphics[scale=.435]{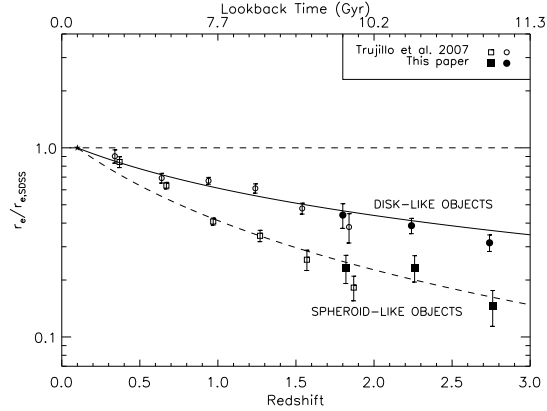}
\caption{Size evolution of massive galaxies ($M > 10^{11} M_{\odot}$) with redshift. Plotted is the ratio of the median sizes of galaxies in our sample 
with respect to sizes of nearby galaxies in the SDSS ~\cite{3} local comparison (solid points). The results of ~\cite{13} for systems at $0.2<z<2$ are overplotted (open symbols). The error bars indicate the uncertainty ($1\sigma$) at the median position.}
\label{fig:2}       
\end{figure}

To quantify  the observed size evolution we show the ratio between the sizes we measure, and the measured sizes of nearby galaxies at the same
mass, as a function of redshift (Fig. ~\ref{fig:2}), using again the SDSS ~\cite{3} as the local sample. We fit the evolution of the decrease in half-light ratio with redshift as a power-law $\sim \alpha(1+z)^{\beta}$, where we calculate that for the disk-like galaxies $\beta = -0.82\pm0.03$, and for the spheroid-like systems $\beta = -1.49\pm0.04$.

%
\section{Discussion}
\label{sec:4}
As has been demonstrated in previous work, many massive galaxies at $z < 2$ appear to grow in size by up to an order of magnitude (e.g., ~\cite{13}).  An interesting question is whether these massive galaxies become progressive smaller at higher redshifts, containing possibly even smaller sizes at $z>2$.  Our results provide the first statistical sample in which to answer this question.  As seen in Fig. 2, the objects in our sample are compatible with the idea that the size evolution reaches a plateau beyond $z=2$. In this plot, the significance of evolving to smaller sizes is $2.2\sigma$ for disks, and $1.8\sigma$ for spheroids. To shed some light on this question we compute the stellar mass density of our galaxies and compare these to the densest collection of stars in the local Universe -- globular clusters.

A typical spheroid--like galaxy in our sample at z$\sim$2.75 has a stellar mass of $\sim$2$\times$10$^{11}$M$_{\odot}$, and a size of r$_e$$\sim$1 kpc. The stellar density for this object, assuming spherical symmetry, is $\rho=(0.5M)/(4/3\pi r_{e}^{3})$$\sim$2.4$\times$10$^{10}$M$_{\odot}{\rm kpc}^{-3}$. A disk--like galaxy 
at z$\sim$2.75 has a typical mass of $\sim$2$\times$10$^{11}$M$_{\odot}$ and size r$_e$$\sim$2 kpc. Assuming a disk symmetry, the stellar mass density within 
these disk-like systems is $\rho=(0.5M)/(\pi r_{e}^{2}h)$$\sim$2.6$\times$10$^{10}$M$_{\odot}{\rm kpc}^{-3}$, where we have used h$\sim$0.3 kpc.
In both cases the stellar mass densities are similar. A typical globular cluster (r$_e$=10 pc and M$\simeq$10$^5$M$_{\odot}$) has a density of
$\sim$1.2$\times$10$^{10}$M$_{\odot}$kpc$^{-3}$. This is remarkably similar to our massive galaxies at $z > 2$, and reveals that these high--z galaxies may in principle have an origin similar to globular clusters. These high densities also suggest that their stellar mass densities likely do not become much larger at high redshifts ($z > 3$). A massive galaxy at $z > 2$ must have formed very quickly, and these high stellar densities could reflect the high gas densities in the primeval Universe.

%

%
%

\begin{thebibliography}{99.}%
%
%
\bibitem{1} Daddi, E., et al. 2005, ApJ, 626, 680
\bibitem{2} Trujillo I., Feulner G., Goranova Y. et al. 2006b, MNRAS, 373, L36
\bibitem{3} Shen et al., 2003, MNRAS, 343, 978
\bibitem{4} Khochfar S. \& Silk J. 2006, ApJ, 648, L21
\bibitem{5} Boylan-Kolchin M., Ma C-P., Quataert E., 2006, MNRAS, 369, 1081
\bibitem{6} van Dokkum, P. G. et al. 2008, ApJ, 677, L5
\bibitem{7} Buitrago, F. et al. 2008, ApJ, 687, L61
\bibitem{8} Giavalisco M., et al., 2004, ApJ, 600, L93
\bibitem{9} Conselice, C.J. et al. 2007, MNRAS, 381, 962
\bibitem{10} Bruzual G. \& Charlot S., 2003, MNRAS, 344, 1000
\bibitem{11} Peng C. Y., Ho L. C., Impey C. D., Rix H. W., 2002, AJ, 124, 266
\bibitem{12} Trujillo I., F$\ddot{o}$rster Schreiber N. M., Rudnick G., et al., 2006, ApJ, 650, 18
\bibitem{13} Trujillo I., Conselice C. J., Bundy K., Cooper M. C., Eisenhardt P., Ellis R., 2007, MNRAS, 382, 109
%
\end{thebibliography}
\end{document}